\documentclass[showpacs,aps,preprint]{revtex4}
\usepackage{graphicx}
\usepackage{amsmath,amssymb}

\def\gtsim{\mathrel{\hbox{\raise0.2ex
  \hbox{$>$}\kern-0.75em\raise-0.9ex\hbox{$\sim$}}}}
\def\ltsim{\mathrel{\hbox{\raise0.2ex
  \hbox{$<$}\kern-0.75em\raise-0.9ex\hbox{$\sim$}}}}

\begin{document}

\title{
Realistic description of the rotational bands in rare earth nuclei by
angular-momentum-projected multi-cranked configuration-mixing method.}

\author{Mitsuhiro Shimada, Shingo Tagami and Yoshifumi R. Shimizu}
\affiliation{Department of Physics, Graduate School of Science,
Kyushu University, Fukuoka 819-0395, Japan}


\begin{abstract}

Recently we have proposed a reliable method to describe
the rotational band in a fully microscopic manner.
The method has recourse to the configuration-mixing
of several cranked mean-field wave functions
after the angular-momentum-projection.
By applying the method with the Gogny D1S force as an effective interaction,
we investigate the moments of inertia of the ground state rotational bands
in a number of selected nuclei in the rare earth region.
As another application we try to describe, for the first time,
the two-neutron aligned band in $^{164}$Er,
which crosses the ground state band
and becomes the yrast states at higher spins.
Fairly good overall agreements with the experimental data are achieved;
for nuclei, where the pairing correlations are properly described,
the agreements are excellent.
This confirms that the previously proposed method is
really useful for study of the nuclear rotational motion.

\end{abstract}

\pacs{21.10.Re, 21.60.Ev, 23.20.Lv}

\maketitle

\section{Introduction}
\label{sec:intro}

The nuclear rotation is one of the most typical collective motions
in nuclei~\cite{BM75}.  Its semiclassical nature makes it possible
to introduce the classical concepts like the rotational frequency
and the transformation to the rotating frame, which are very useful
to analyze the high-spin rotational bands~\cite{BF79} and nowadays
provide a standard method called the cranked shell model~\cite{BFM86}.
However, the nucleus is a quantum many-body system and the basis of
such a semiclassical treatment of the rotational motion is
the symmetry breaking caused by the deformed mean-field, see e.g.~\cite{RS80}.
In fact, the rotational motion emerges as a symmetry restoring
collective motion of atomic nucleus as a whole and can be described
full-quantum mechanically by the angular-momentum-projection method~\cite{RS80}.
Although a nice rotational spectrum can be obtained by
the angular-momentum-projection from the deformed mean-field state,
it has been known that the level spacing of the obtained rotational spectrum
tends to be larger than that of the experimental data;
i.e. the moment of inertia is quite often underestimated.
Inclusion of the time-odd components into the deformed mean-field,
from which the projection performed, improves this problem and
it can be easily realized by the cranking procedure~\cite{TS12}.

It has been demonstrated that the small cranking frequency is
enough to increase the moment of inertia and the result of projection
does not depend on the actual value of the frequency; we call it
``infinitesimal cranking'' for the angular-momentum-projection~\cite{TS16}.
Recently we have extended the study of the rotational motion by
the angular-momentum-projection method~\cite{STS15}.  Namely the cranking
procedure is combined with the projection by employing
the configuration-mixing with respect to the finite rotational frequency;
we call it angular-momentum-projected
multi-cranked configuration-mixing.  This method was originally proposed
by Peierls and Thouless long time ago~\cite{PT62}, but has not been
taken seriously.  We have applied it to a few examples to show that
it gives a reliable description of the rotational motion
at high-spin states~\cite{STS15}.
The angular-momentum-projected configuration-mixing with respect to
a few cranked mean-field states has recently been performed
also in Ref.~\cite{RER15}.

As for the application of the angular-momentum-projection method
to the nuclear rotational motion, many pioneering works have been done
by the Projected Shell Model, see e.g. Ref.~\cite{HS95}.
While the basic idea is the same, much larger but simple configurations,
like the zero-, two-, four-, ..., quasiparticle excited bands are mixed
in the sense of the shell model.
We believe the multi-cranked configuration-mixing~\cite{STS15}
is an alternative, which incorporates a relatively small number of
mean-field configurations with the help of the cranking procedure.

The main purpose of the present work is to demonstrate that the multi-cranked
configuration-mixing is indeed a reliable method to describe
the rotational band with the angular-momentum-projection method.
We first apply the method to the ground-state rotational bands for
a number of selected nuclei in the rare earth region.
At high-spin states, it is well-known that
the band crossing (back-bending) phenomenon
between the ground-state (g-) band and the Stockholm (s-) band,
i.e., the two-neutron aligned band, occurs.
Therefore, we try to study the s-band in a typical nucleus $^{164}$Er
with the same multi-cranked configuration-mixing method;
we are able to study the g- and s-bands separately
without the inter-band mixing.
The cranked mean-field states are determined selfconsistently by
the Hartree-Fock-Bogoliubov (HFB) method for given rotational frequencies
employing the finite-range Gogny interaction~\cite{DeGo80} with
the D1S parameter set~\cite{D1S}.  After briefly explaining the theoretical
framework in Sec.~\ref{sec:multi}, we show the results of calculations
in Sec.~\ref{sec:results}.  The conclusion is drawn in Sec.~\ref{sec:concls}

\section{Theoretical framework}
\label{sec:multi}

Our basic approach to study the high-spin states of
the nuclear collective rotation
is the angular-momentum-projected configuration-mixing,
or the projected generator coordinate method (GCM),
where the cranking frequency $\omega_{\rm rot}$ is
employed as a generator coordinate.
It was first proposed by Peierls-Thouless~\cite{PT62},
and the wave function is calculated by
\begin{equation}
 |\Psi^I_{M,\alpha}\rangle =
 \int d\omega_{\rm rot}\sum_{K} g^I_{K,\alpha}(\omega_{\rm rot})\,
 \hat P^I_{MK}|\Phi_{\rm cr}(\omega_{\rm rot})\rangle,
\label{eq:PTanz}
\end{equation}
where the operator $\hat P^I_{MK}$ is the angular momentum projector,
and the mean-field wave function, $|\Phi_{\rm cr}(\omega_{\rm rot})\rangle$,
is obtained by the selfconsistent cranking procedure
with the cranked Hamiltonian, $H-\omega_{\rm rot}J_y$,
\begin{equation}
\delta\langle\Phi_{\rm cr}(\omega_{\rm rot})|
 H-\omega_{\rm rot}J_y |\Phi_{\rm cr}(\omega_{\rm rot})\rangle=0.
\label{eq:crank}
\end{equation}
In the present work, the ground-state mean-field states are axially deformed
and the cranking axis is chosen to be the $y$-axis perpendicular
to the symmetry axis ($z$-axis).
Practically we discretize the generator coordinate, i.e., the cranking frequency,
as ($\omega^{(n)}_{\rm rot}$; $n=1,2,\cdots,n_{\rm max}$)
in Eq.~(\ref{eq:PTanz}),
\begin{equation}
 |\Psi^I_{M,\alpha}\rangle = \sum_{Kn} g^I_{Kn,\alpha}\,
 \hat P^I_{MK}|\Phi_{\rm cr}(\omega_{\rm rot}^{(n)})\rangle,
\label{eq:proj}
\end{equation}
and solve the configuration-mixing amplitude,
$g^I_{Kn,\alpha}=g^I_{K,\alpha}(\omega_{\rm rot}^{(n)})$,
with the so-called Hill-Wheeler equation,
\begin{equation}
 \sum_{K'n'}{\cal H}^I_{Kn,K'n'}\ g^I_{K'n',\alpha} =
 E^I_\alpha\,
 \sum_{K'n'}{\cal N}^I_{Kn,K'n'}\ g^I_{K'n',\alpha},
\label{eq:HW}
\end{equation}
where the Hamiltonian and norm kernels are defined as usual,
\begin{equation}
 \left\{ \begin{array}{c}
   {\cal H}^I_{Kn,K'n'} \\ {\cal N}^I_{Kn,K'n'} \end{array}
 \right\} = \langle \Phi_{\rm cr}(\omega_{\rm rot}^{(n)}) |
 \left\{ \begin{array}{c}
   H \\ 1 \end{array}
 \right\} \hat{P}_{KK'}^I | \Phi_{\rm cr}(\omega_{\rm rot}^{(n')})\rangle.
\label{eq:kernels}
\end{equation}
We do not perform the number projection in the present work,
and treat the number conservation approximately
by replacing $H \rightarrow H-\lambda_\nu (N-N_0)-\lambda_\pi (Z-Z_0)$,
where $N_0$ and $Z_0$ are the neutron and proton numbers to be fixed.
As for the neutron and proton chemical potentials
$\lambda_\nu$ and $\lambda_\pi$
we use those obtained for the HFB ground-state.

We have recently developed an efficient method for
the angular-momentum-projection and the configuration-mixing~\cite{TS12}.
This method is fully utilized also in the present work.
To solve the HFB equation and to perform the projection calculation 
the harmonic oscillator basis expansion is employed.
More details of our theoretical framework can be found
in Refs.~\cite{TS12,STS15,TS16}.

\section{Results of calculation}
\label{sec:results}

\subsection{Ground-state bands of rare earth nuclei}
\label{sec:gr-rea}

As it is mentioned we employ the Gogny force
with the D1S parameter set~\cite{D1S} as an effective interaction.
Therefore there is no ambiguity for the Hamiltonian.
The HFB equation is solved in the space generated
by the isotropic harmonic oscillator potential
with the frequency $\hbar\omega=41/A^{1/3}$ MeV.
The size of the space is controlled by the oscillator quantum number
$N_{\rm osc}^{\rm max}$; all the basis states satisfying
$n_x+n_y+n_z \le N_{\rm osc}^{\rm max}$ are included.
We use $N_{\rm osc}^{\rm max}=10$ for the following systematic calculations
of the rotational spectra.
The main target of the present work is the most basic rotational band,
i.e., the g-band.  Therefore,
we selected typical deformed nuclei in the rare earth region, i.e.,
three isotopes with the neutron number in $N=92$--100
in each Gd, Dy, Er, and Yb nuclide,
as they are tabulated in Table~\ref{tab:mf_rare}.
In this table we show the nuclear radii, the deformation parameters,
and the average pairing gaps obtained by the HFB calculations
in the ground-states of these selected nuclei.
The ground-states are axially symmetric in all nuclei
and the deformation parameter $\beta_\lambda$ is defined by~\cite{IMYM02}
\begin{equation}
 \beta_\lambda=\frac{4\pi}{3}\frac{Q_{\lambda 0}}{A R^\lambda},
\label{eq:def}
\end{equation}
where the $\lambda$-pole moment $Q_{\lambda 0}$ and the radius $R$ 
are calculated by the expectation value with respect to the HFB state,
\begin{equation}
 Q_{\lambda 0}=\biggl\langle
 \sum_{i=1}^A (r^\lambda Y_{\lambda 0})_i \biggr\rangle,
 \qquad
 R=\left[\frac{5}{3A}\biggl\langle
 \sum_{i=1}^A (r^2)_i \biggr\rangle\right]^{1/2}.
\label{eq:Qmom}
\end{equation}
The average pairing gap is defined by
\begin{equation}
\bar{\Delta}=
-\left[\sum_{a>b}\Delta_{ab}\kappa^*_{ab}\right]
 \left[\sum_{a>0}\kappa^*_{a\tilde{a}}\right]^{-1},\qquad
  \Delta_{ab}= \sum_{c>d}\bar{v}_{ab,cd}\,\kappa_{cd},
\label{eq:avgap}
\end{equation}
where the quantities $\bar{v}_{ab,cd}$ and $\kappa_{ab}$
are the anti-symmetrized matrix element of the two-body interaction
and the abnormal density matrix (pairing tensor), respectively~\cite{RS80},
and $\tilde{a}$ means the time-reversal conjugate state of $a$.
Here we corrected the misprinted expression of denominator
in the definition of $\bar{\Delta}$ in Refs.\cite{STS15,TS16}.
The even-odd mass differences calculated
by the 4th-order difference formula based on the 2003 mass table~\cite{AW03}
are also included in Table~\ref{tab:mf_rare}.

\begin{table*}[!htb]
\begin{center}
\begin{tabular}{c p{1pt} c p{1pt} c p{1pt} c p{1pt} c c p{1pt} c c p{1pt} c}
\hline \hline
  &&  &&  &&    && \multicolumn{2}{c}{$\bar{\Delta}$ [MeV]}
      && \multicolumn{2}{c}{$\Delta_{\mbox{\scriptsize e-o}}$ [MeV]} \\
\cline{8-10} \cline{12-13}
 nuclide && $R$ [fm] && $\beta_2$ && $\beta_4$
 && neutron & proton && neutron & proton && norm cut-off \\
\hline
$^{156}$Gd$_{92}$ && 6.64 && 0.308 && 0.160 && 0.649 & 0.979 && 1.004 & 0.968 && $10^{-11}$ \\
$^{158}$Gd$_{94}$ && 6.67 && 0.319 && 0.150 && 0.704 & 0.948 && 0.884 & 0.901 && $10^{-11}$ \\
$^{160}$Gd$_{96}$ && 6.71 && 0.327 && 0.135 && 0.739 & 0.922 && 0.794 & 0.875 && $10^{-11}$ \\
$^{158}$Dy$_{92}$ && 6.66 && 0.304 && 0.133 && 0.778 & 0.946 && 1.034 & 1.081 && $10^{-12}$ \\
$^{162}$Dy$_{96}$ && 6.73 && 0.323 && 0.112 && 0.791 & 0.830 && 0.873 & 0.951 && $10^{-12}$ \\
$^{164}$Dy$_{98}$ && 6.76 && 0.328 && 0.096 && 0.714 & 0.799 && 0.825 & 0.879 && $10^{-11}$ \\
$^{160}$Er$_{92}$ && 6.67 && 0.281 && 0.108 && 0.838 & 1.027 && 1.112 & 1.207 && $10^{-11}$ \\
$^{162}$Er$_{94}$ && 6.72 && 0.305 && 0.105 && 0.847 & 0.934 && 1.066 & 1.125 && $10^{-12}$ \\
$^{164}$Er$_{96}$ && 6.75 && 0.316 && 0.090 && 0.845 & 0.858 && 1.020 & 1.025 && $10^{-11}$ \\
$^{164}$Yb$_{94}$ && 6.73 && 0.282 && 0.088 && 0.877 & 1.026 && 1.148 & 1.203 && $10^{-10}$ \\
$^{168}$Yb$_{98}$ && 6.81 && 0.321 && 0.064 && 0.783 & 0.822 && 0.993 & 1.017 && $10^{-11}$ \\
$^{170}$Yb$_{100}$&& 6.84 && 0.325 && 0.046 && 0.615 & 0.682 && 0.840 & 0.945 && $10^{-11}$ \\
\hline \hline
\end{tabular}
\vspace*{4mm}
\caption{
Nuclear radii, deformation parameters $\beta_2$ and $\beta_4$,
and the average pairing gaps $\bar{\Delta}$
for neutrons and protons obtained by the non-cranked
($\omega_{\rm rot} = 0$) HFB calculation with $N_{\rm osc}^{\rm max}=10$.
Experimental even-odd mass differences calculated by the 4th-order formula
$\Delta_{\mbox{\scriptsize e-o}}$ and the value of the norm cut-off
for the configuration-mixing calculation are also included.
}
\label{tab:mf_rare}
\end{center}
\end{table*}

The result of the deformation parameter $\beta_2$ roughly corresponds
to experimental data deduced by the measured $B(E2)$ values~\cite{NNDC},
but the calculated $\beta_2$ are slightly smaller.
This is merely due to the differences in the definitions
of $\beta_2$ in Ref.~\cite{NNDC} and in the present work.
The calculated radius $R$ is 2.4--2.9\% larger than the empirical value
$1.2A^{1/3}$ [fm], which is mostly due to the effect of deformation.
In these stable nuclei the difference of the radius
from the empirical value is not so large,
but for the unstable nuclei the difference is expected to be larger.
Therefore, it is important to measure
the nuclear radius to reliably extract
the deformation parameters~\cite{Shimada16}.

As for the average pairing gaps selfconsistently calculated values are
smaller than the even-odd mass differences in most cases.
Especially the neutron pairing gap in $^{156}$Gd is only 65\% of
the even-odd mass difference, and in $^{170}$Yb the calculated average gaps
for both the neutron and proton are about 27\% smaller.  On the other hand,
for $^{160}$Gd, $^{162}$Dy and $^{164}$Dy, the calculated gaps relatively
well correspond to the even-odd mass differences, and their differences
are less than 14\%.  As it is discussed in the following, the agreement
of the calculated gaps with the even-odd mass differences is crucial
to reproduce the moments of inertia of the g-band.

If the deformed superconducting state is obtained for the ground-state,
we generate the cranked HFB states in Eq.~(\ref{eq:crank})
for a given set of rotational frequencies
($\omega^{(n)}_{\rm rot}$; $n=1,2,\cdots,n_{\rm max}$).
It was demonstrated~\cite{STS15} that the result of configuration-mixing
does not depend on the choice of a set of frequencies,
if the number of frequencies, $n_{\rm max}$, is five.
Therefore, we take $n_{\rm max}=5$ and choose them (almost) equidistantly.
Since the ground-state rotational band is studied, only the frequencies
before the g-s crossing should be selected; we choose
$\hbar\omega^{(n)}_{\rm rot}=0.01,\,0.05,\,0.10,\,0.15,\,0.20$ MeV
as in Ref.~\cite{STS15}, for most of nuclei in Table~\ref{tab:mf_rare}.
However, it is known that
there is no sharp g-s crossing observed in N=98 isotopes, so that we choose
$\hbar\omega^{(n)}_{\rm rot}=0.01,\,0.075,\,0.150,\,0.225,\,0.300$ MeV
for $^{164}$Dy and $^{168}$Yb.

Once the cranked HFB states are prepared we perform
the angular-momentum-projected multi-cranked configuration-mixing,
see Eq.~(\ref{eq:proj}), to obtain the rotational spectrum.
In order for the efficient calculation of projection,
the cut-off of the quasiparticle basis is employed~\cite{TS12},
namely the canonical basis states of the HFB wave function
whose occupation numbers are larger than $10^{-6}$ are only retained.
As for the integration mesh points for the Euler angles
($\alpha,\beta,\gamma$) in the angular-momentum-projector,
we take $N_\alpha=N_\gamma=2K_{\rm max}+2$ and $N_\beta=2I_{\rm max}+2$.
For the present systematic calculation of the ground-state band,
$K_{\rm max}=16$ and $I_{\rm max}=30$ are chosen; the smaller value
of $K_{\rm max}$ is enough because the HFB wave function is nearly
axially symmetric ($K$ mixing is mainly induced by the cranking procedure).

In the configuration-mixing in Eq.~(\ref{eq:proj}),
the superposition of the states with respect to $(K,n)$ for given $I$
is overcomplete and there are vanishingly small norm states,
which causes numerical problems~\cite{RS80}.
Therefore, the norm cut-off should be done;
namely, the eigenvalues of the norm kernel are first calculated
and the small norm states should be excluded
when solving the Hill-Wheeler Eq.~(\ref{eq:HW}).
The value for the norm cut-off is better to be as small as possible
not to miss important contributions.
We start from the value  $10^{-12}$ and increase it
to avoid the numerical problems in each case.
The actual values used in the following calculation are also denoted
in Table~\ref{tab:mf_rare}; they are in the range, $10^{-10}-10^{-12}$.

\begin{figure*}[!htb]
\begin{center}
\includegraphics[width=140mm]{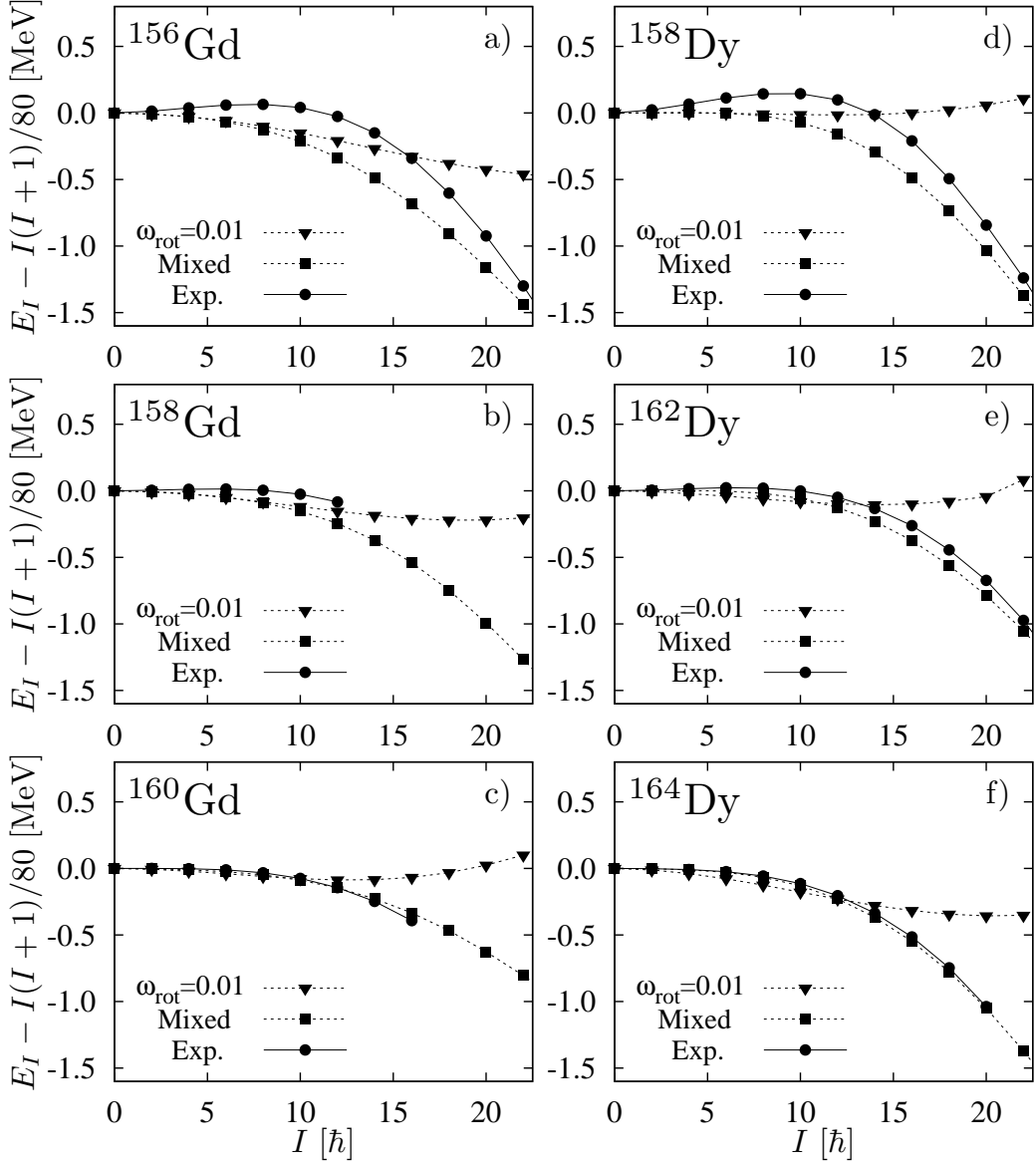}
\vspace*{-4mm}
\caption{
Rotational energy spectra for the g-band
of the selected Gd isotopes, a)--c), and Dy isotopes, d)--f).
The reference rotational energy, $I(I+1)/80$~MeV, is subtracted.
The result of simple projection from one cranked HFB state
with $\hbar\omega_{\rm rot}=0.01$~MeV as well as that
of the projected multi-cranked configuration-mixing (Mixed)
are included in addition to the experimental data (Exp.).
}
\label{fig:EIr1}
\end{center}
\end{figure*}

\begin{figure*}[!htb]
\begin{center}
\includegraphics[width=140mm]{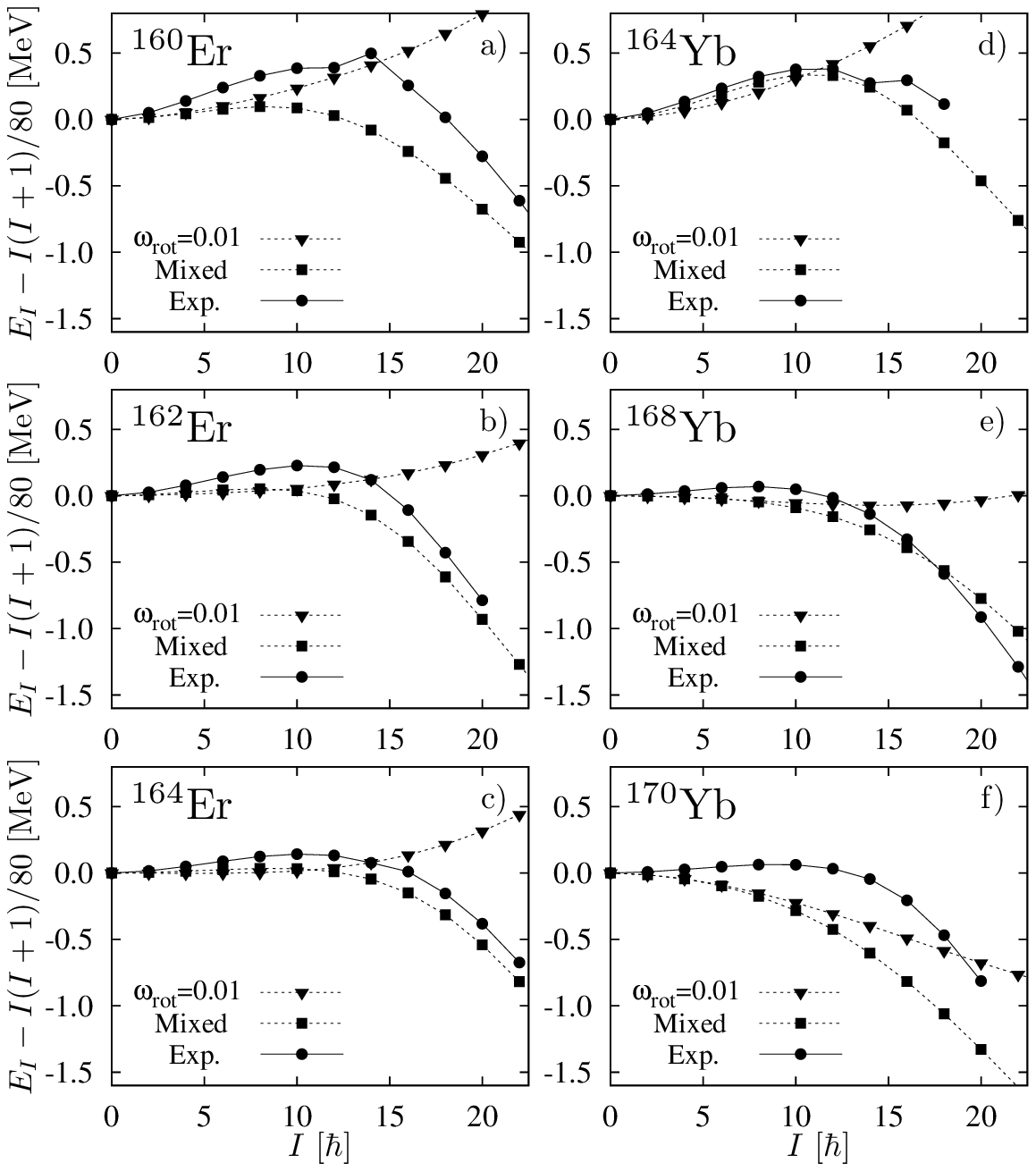}
\vspace*{-4mm}
\caption{
The same as Fig.~\ref{fig:EIr1} but for the selected
Er isotopes, a)--c), and Yb isotopes, d)--f).
}
\label{fig:EIr2}
\end{center}
\end{figure*}

Figures~\ref{fig:EIr1} and~\ref{fig:EIr2} show the resultant
rotational spectra of the configuration-mixing
for the ground-state bands in nuclei in Table~\ref{tab:mf_rare}.
The results of the simple projection from one cranked HFB state
with small frequency $\hbar\omega_{\rm rot}=0.01$~MeV,
i.e., those of the infinitesimal cranking, are also included.
The experimental data are taken from the Table of Isotope homepage~\cite{TOI}.
To show the detail we subtract the reference rotational energy,
$I(I+1)/80$ MeV, in each spectrum.
By comparing the results of the configuration-mixing and
of the simple projection from one infinitesimally cranked HFB state
in these figures, it is clear that the infinitesimal cranking gives
a good description of the low-spin states.
However, the deviations become non-negligible quickly at higher spins,
$I \gtsim 10\,\hbar$.
With the multi-cranked configuration-mixing the energy gain at $I=20\,\hbar$
from the simple projection is about $0.7-1.5$ MeV, and therefore the effect of
configuration-mixing is crucial for the description of the high-spin states.

In comparison with the experimental spectra, the results of
the configuration-mixing reproduce the bending-down behaviors of the data
at higher spins, which reflects the increase of moments of inertia.
In contrast, the spectra of the simple projection
with the infinitesimal cranking do not change
or even increase as spin increases, which means that
the moments of inertia obtained by the simple projection are rather constant.
The deviation of the calculated spectrum from the measured spectrum
is rather large in $^{156}$Gd, $^{158}$Dy, $^{160}$Er, and $^{170}$Yb,
for which the average pairing gaps are considerably smaller than
the even-odd mass differences,
especially for neutron, see Table~\ref{tab:mf_rare}.
On the other hand, the agreements are almost perfect for
$^{160}$Gd, $^{162}$Dy, $^{164}$Dy, and $^{164}$Yb,
in which the calculated average pairing gaps for both neutron and proton
well correspond to the even-odd mass differences.  Thus the reproduction
of the pairing properties is very important to achieve a good description
of the ground-state rotational band, which is a rather well-known fact.
It should be emphasized that
the agreements for other nuclei are rather satisfactory considering
the fact that we have no room for adjustment in the present calculations.
The relative difference between the isotopes, e.g.,
between $^{162}$Er and $^{164}$Er, or between $^{164}$Yb and $^{168}$Yb,
is also reproduced.

\begin{figure*}[!htb]
\begin{center}
\includegraphics[width=140mm]{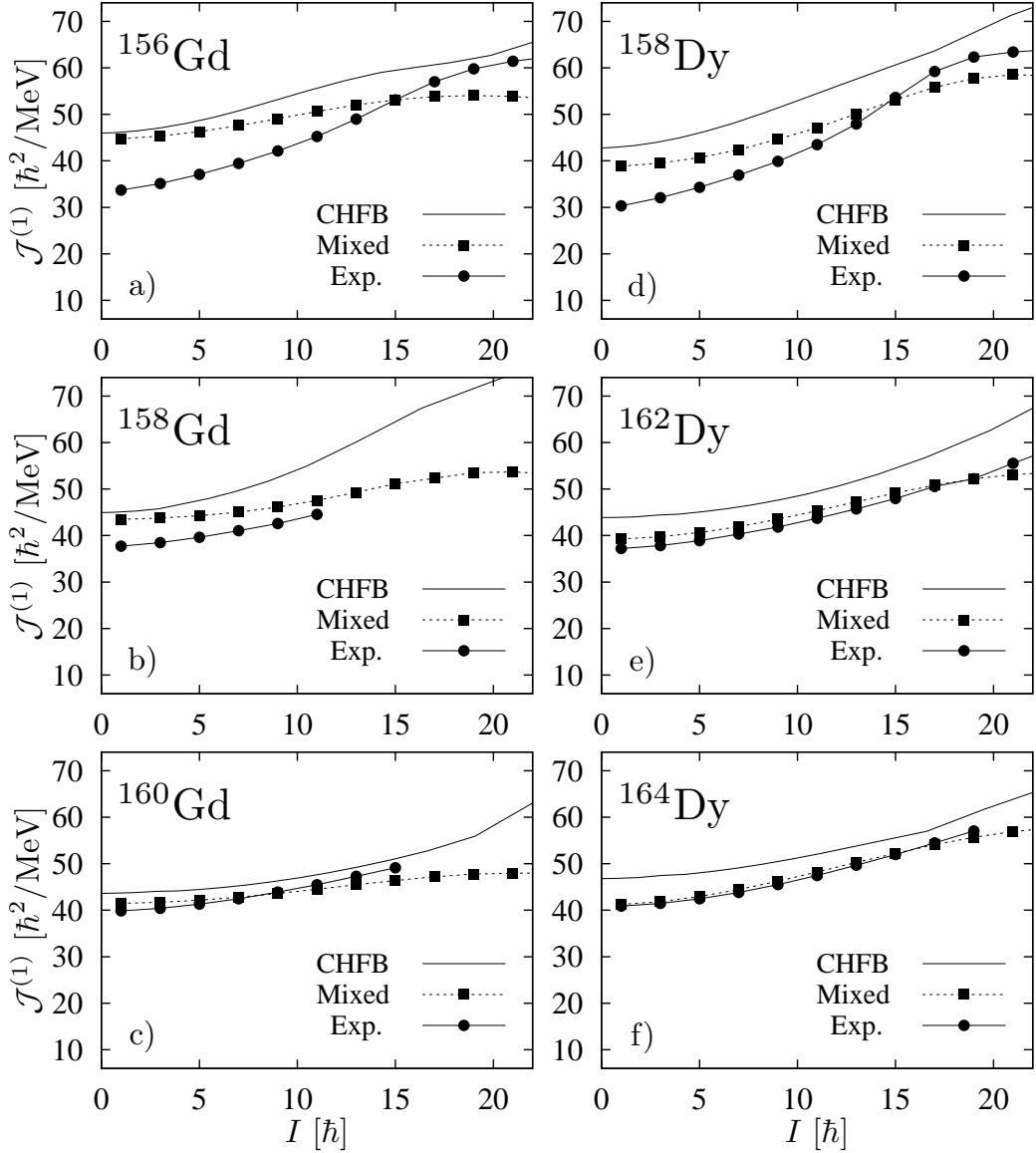}
\vspace*{-4mm}
\caption{
Moments of inertia versus spin value for the g-band
of the selected Gd isotopes, a)--c), and Dy isotopes, d)--f).
The results of the cranked HFB (CHFB) and the configuration-mixing (Mixed)
are compared with the experimental data (Exp.);
see the text for the precise definition of
the first moment of inertia ${\cal J}^{(1)}$.
}
\label{fig:EMo1}
\end{center}
\end{figure*}

\begin{figure*}[!htb]
\begin{center}
\includegraphics[width=140mm]{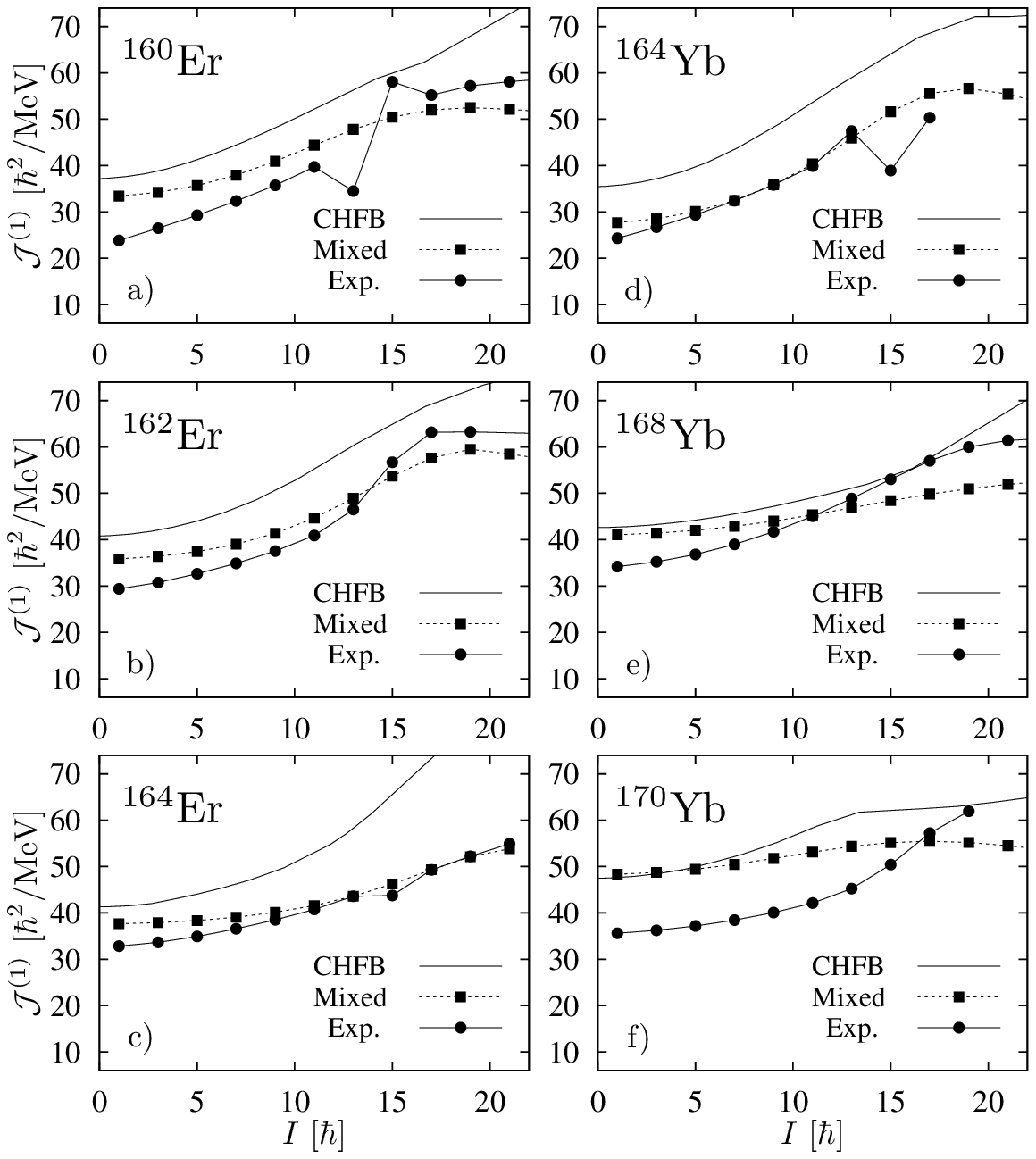}
\vspace*{-4mm}
\caption{
The same as Fig.~\ref{fig:EMo1} but for the selected
Er isotopes, a)--c), and Yb isotopes, d)--f).
}
\label{fig:EMo2}
\end{center}
\end{figure*}

To study the rotational property in more detail, we show the first
(or kinematic) moment of inertia in Figs.~\ref{fig:EMo1} and~\ref{fig:EMo2},
which is defined by
\begin{equation}
 {\cal J}^{(1)}(I) = \frac{(2I+1)\hbar^2}{E(I+1)-E(I-1)}.
\label{eq:J1}
\end{equation}
As a reference, the result of the cranked HFB calculation
is also included in these figures, which is calculated by
\begin{equation}
 {\cal J}^{(1)}(\omega_{\rm rot})= \frac{\langle\Phi_{\rm cr}(\omega_{\rm rot})|
 J_y|\Phi_{\rm cr}(\omega_{\rm rot})\rangle}{\omega_{\rm rot}},
\label{eq:J1CHFB}
\end{equation}
and is plotted as a function of
\begin{equation}
 I\,\hbar =\langle\Phi_{\rm cr}(\omega_{\rm rot})|
J_y|\Phi_{\rm cr}(\omega_{\rm rot})\rangle - \frac{1}{2}\hbar.
\label{eq:IavCHFB}
\end{equation}
We do not try to search the minimum energy at the fixed spin value
in the cranked HFB calculation, and therefore
the back-bending behavior of the moment of inertia is not obtained.
It should be noticed that the cranked HFB inertia in Eq.~(\ref{eq:J1CHFB})
at high-spin states after the alignment of two quasineutrons
should be considered to be that of the s-band
and the inertia is unphysical in the band crossing region.
The irregularities seen in the experimental moment of inertia
are due to the effect of the g-s band crossing.

It can be seen in Figs.~\ref{fig:EMo1} and~\ref{fig:EMo2} that
the moments of inertia for the ground-state band increase gradually
as functions of spin.  This behavior is quite nicely reproduced
by the multi-cranked configuration-mixing calculations.
The values of moment of inertia are considerably overestimated
at low-spins in $^{156}$Gd, $^{158}$Dy, $^{160}$Er, and $^{170}$Yb;
again this is mainly because the calculated pairing gaps are
markedly smaller than the even-odd mass differences.
In these nuclei the amounts of increase for the moment of inertia
are also much smaller than the experimental data.
This indicates that the increase of the moment of inertia is mainly
related to the reduction of the pairing correlations at higher spins.
The agreements between the calculated and measured inertias
are excellent for $^{160}$Gd, $^{162}$Dy, $^{164}$Dy, and $^{164}$Yb
in the whole spin range shown in the figures.
For the other cases, $^{158}$Gd, $^{162}$Er, $^{164}$Er, and $^{168}$Yb,
the deviations of the calculated moments of inertia from the experimental data
are less than 20\%, which is quite non trivial.
The different spin-dependence observed between the isotopes, e.g.,
between $^{162}$Er and $^{164}$Er, is also nicely reproduced by
the configuration-mixing calculations.
The inertia calculated by the cranked HFB at low-spin
is slightly larger than the result of configuration-mixing;
the large increase of the cranked HFB inertia is caused by
the effect of the two-neutron alignment mentioned above.

It should be emphasized that the moment of inertia obtained by
the simple projection from the non-cranked HFB state
(not shown in the present work) is about 30--40\% smaller than
the result of the infinitesimal cranking~\cite{TS12,STS15,TS16}.
The inclusion of the time-odd components of the wave functions with $K \ne 0$
and the subsequent $K$-mixing is very important to reproduce the correct
magnitude of the moments of inertia.

\begin{figure*}[!htb]
\begin{center}
\includegraphics[width=150mm]{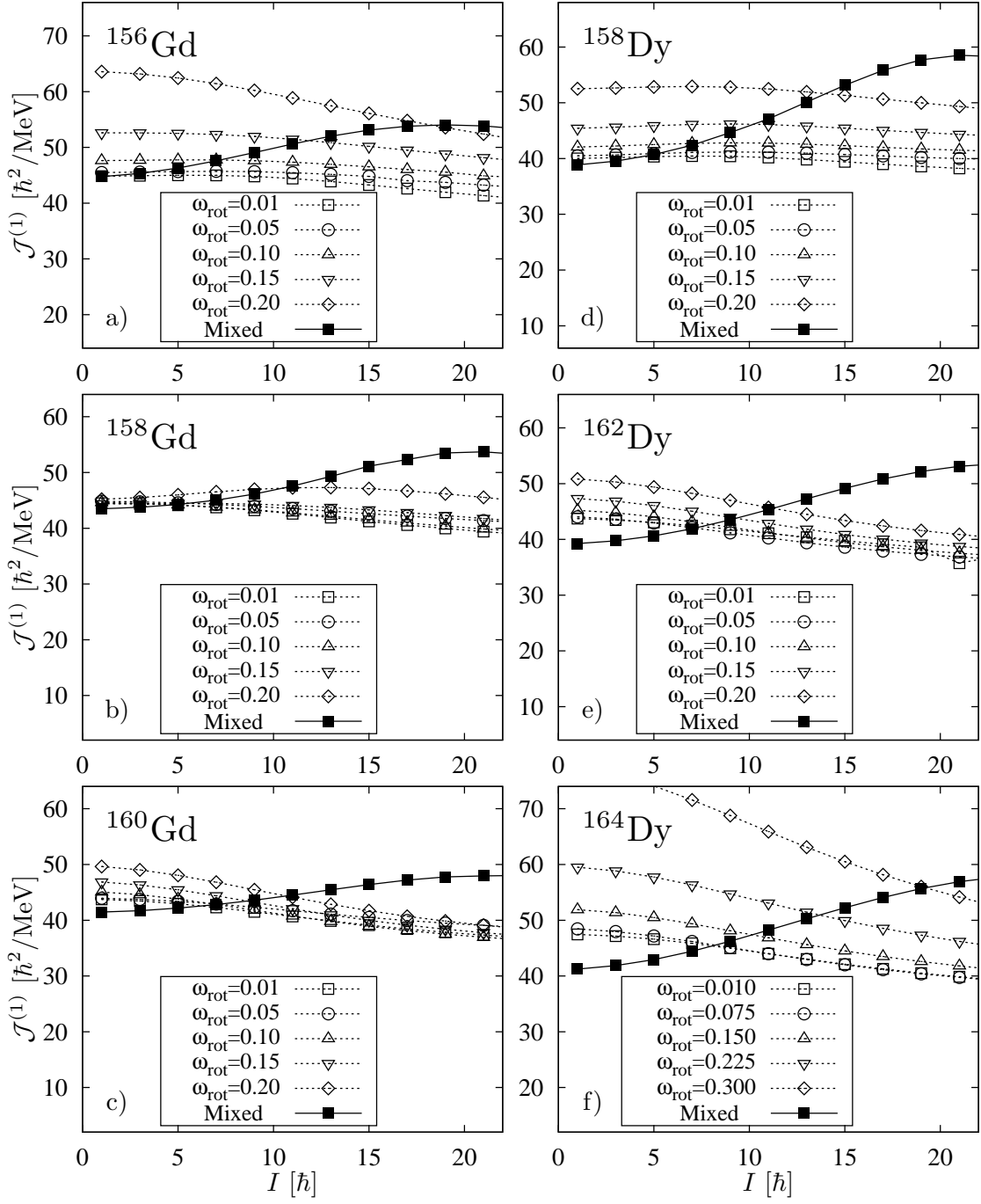}
\vspace*{-4mm}
\caption{
Moments of inertia calculated by the simple projections
from one intrinsic HFB state with five values of the cranking frequency are
compared with the result of the projected configuration-mixing
employing those five HFB states
for the selected Gd isotopes, a)--c), and Dy isotopes, d)--f).
}
\label{fig:EMm1}
\end{center}
\end{figure*}

\begin{figure*}[!htb]
\begin{center}
\includegraphics[width=150mm]{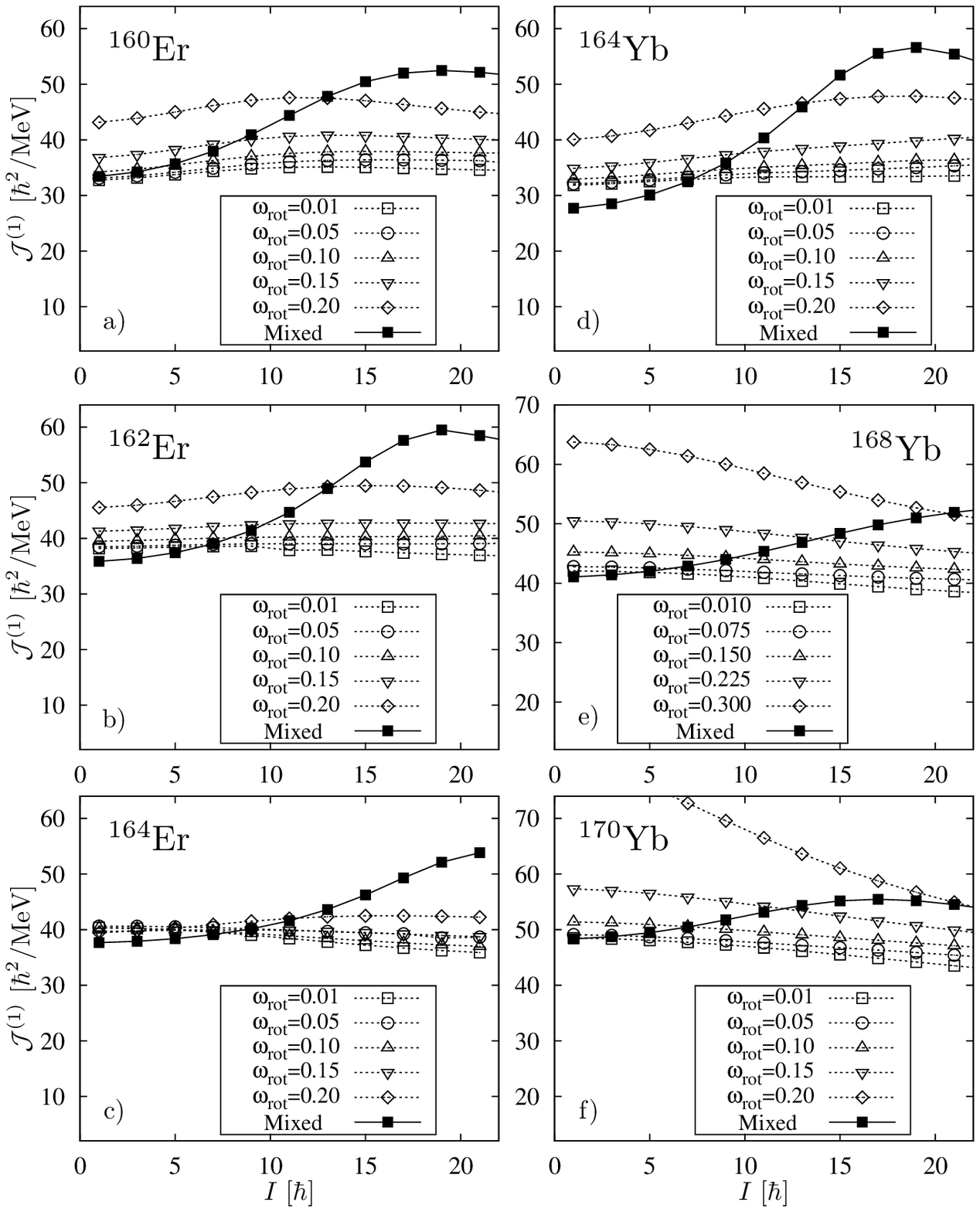}
\vspace*{-4mm}
\caption{
The same as Fig.~\ref{fig:EMm1} but for the selected
Er isotopes, a)--c), and Yb isotopes, d)--f).
}
\label{fig:EMm2}
\end{center}
\end{figure*}

In Figures~\ref{fig:EMm1} and~\ref{fig:EMm2} we show how
the moments of inertia is changed by the multi-cranked configuration-mixing.
Namely, the inertias calculated by
the simple projection from one intrinsic HFB state with five frequencies
are compared with the result of the configuration-mixing.
The calculated inertias with higher cranking frequencies are generally
larger because of the Coriolis anti-pairing effect.
However, those calculated by the simple projection from one HFB states
are almost constant as spin increases or even decrease in several cases.
In contrast, the resultant inertias of the configuration-mixing
always increase as functions of spin in accordance with experimental data.
In this way the configuration-mixing is important to obtain gradually
increasing behavior, which is general for the g-band in the rare earth region.
Note, however, that the amount of increase
is quite different in each nucleus; there is a trend that if the difference
between the inertias calculated with different cranking frequencies
at high-spin is large, then the amount of increase is larger.
It is interesting to note that the behaviors of calculated inertias
by the simple projection with finite frequencies are rather different
in each nucleus.  For example, all five results of the simple projection
are similar in $^{164}$Er, while they are considerably different in $^{170}$Yb.
However, the results of the configuration-mixing make
the behaviors of moment of inertia in all nuclei rather similar, i.e.,
gradually increasing as functions of spin.
It is worthwhile mentioning that the result of the infinitesimal cranking,
i.e., the calculated inertia with $\hbar\omega_{\rm rot}=0.01$ MeV coincides
with that of the configuration-mixing at low-spins in most cases.
This clearly shows that the infinitesimal cranking is enough for
a good description of the rotational band at low-spin states,
although the configuration-mixing is crucial at high-spin states.

\subsection{Description of s-band in $^{164}$Er}
\label{sec:sband}

In the previous section only the g-band is considered.
However, it is well-known that a different rotational band intersects
with the g-band and becomes the yrast state at higher spin values.
This band is called the s-band, in which two quasineutrons are excited
to align their angular momenta to the axis of collective rotation.
The band crossing between the g- and s-bands is the origin of
the back-bending phenomenon first observed in Ref.~\cite{JRH72},
where the rotational frequency decreases when spin increases
along the yrast line.  How this alignment of the two quasineutron
occurs can be nicely understand by the semiclassical cranking model,
see e.g. Refs.~\cite{BFM86}.
We have shown that the g-band can be
nicely described by our multi-cranked configuration-mixing method.
Therefore it is natural to study the s-band with the same method.

\begin{figure}[!htb]
\begin{center}
\includegraphics[width=80mm]{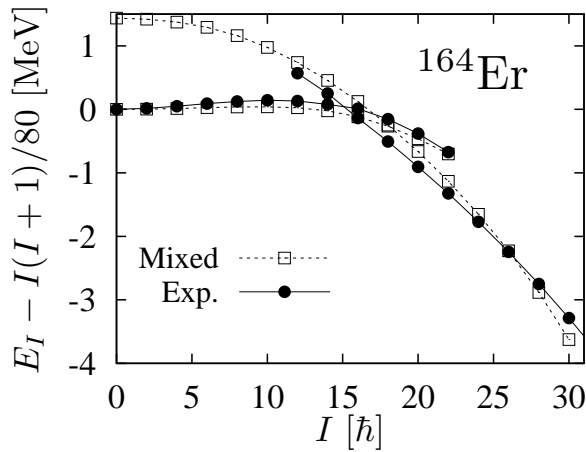}
\vspace*{-4mm}
\caption{
Rotational energy spectra for the g- and s-bands in $^{164}$Er.
The reference rotational energy, $I(I+1)/80$~MeV, is subtracted.
The result of the projected multi-cranked configuration-mixing (Mixed)
is compared with the experimental data (Exp.).
}
\label{fig:Er164gsbr}
\end{center}
\end{figure}

\begin{figure}[!htb]
\begin{center}
\includegraphics[width=80mm]{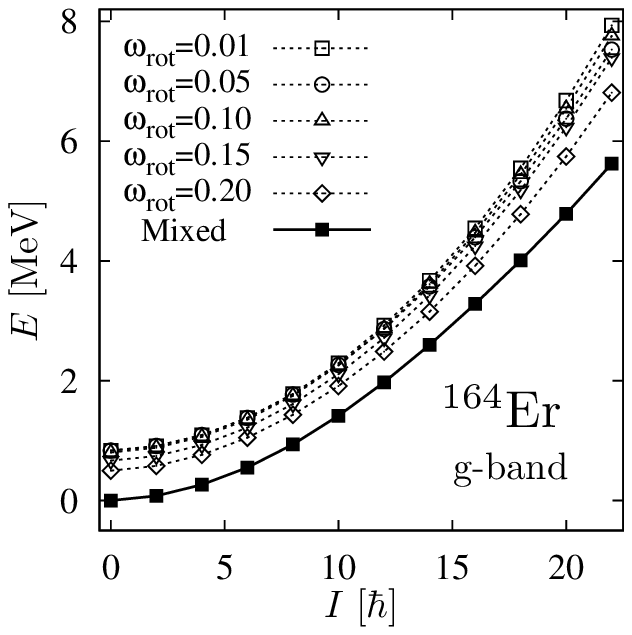}
\vspace*{-4mm}
\caption{
Energy spectra of the g-band in $^{164}$Er obtained by
the simple projection from one intrinsic HFB state
with five values of the cranking frequencies,
$\hbar\omega_{\rm rot}=$0.01, 0.05, 0.10, 0.15, 0.20~MeV,
compared with the result of the projected configuration-mixing
employing those five cranked HFB states.
The energy origin is taken as the energy of the ground state of
the configuration-mixing calculation.
}
\label{fig:Er164gb5m}
\end{center}
\end{figure}

\begin{figure}[!htb]
\begin{center}
\includegraphics[width=80mm]{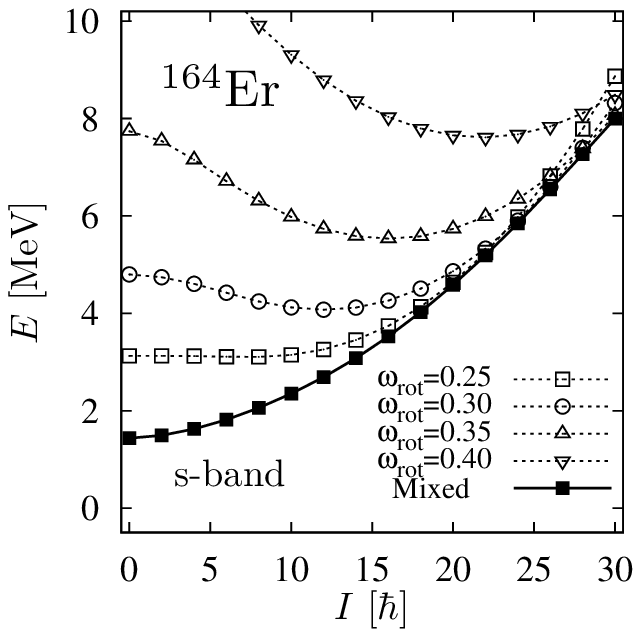}
\vspace*{-4mm}
\caption{
Energy spectra of the s-band in $^{164}$Er obtained by
the simple projection from one intrinsic HFB state
with four values of the cranking frequencies,
$\hbar\omega_{\rm rot}=$0.25, 0.30, 0.35, 0.40~MeV,
compared with the result of the projected configuration-mixing
employing those four cranked HFB states.
The energy origin is taken as the energy of the ground state of
the configuration-mixing calculation.
}
\label{fig:Er164sb4m}
\end{center}
\end{figure}

We have already investigated the g-band in $^{164}$Er
in our previous work~\cite{STS15} with $N_{\rm osc}^{\rm max}=12$.
Therefore, we use the same model space for the s-band in this section.
The method of the calculation is the same as in Ref.~\cite{STS15}
except that the set of cranking frequencies is suitably chosen
for the description of the s-band.
By the cranked HFB calculation, the alignment of the two quasineutrons
occurs at $\hbar\omega_{\rm rot}\approx 0.24$ MeV in $^{164}$Er
with the Gogny D1S parameter set, see Fig.~5 of Ref.~\cite{STS15}.
We use the four cranked HFB states with frequencies,
$\hbar\omega_{\rm rot}=$0.25, 0.30, 0.35, 0.40~MeV,
for the configuration-mixing calculation of the s-band.
The value of the norm cut-off is taken to be $10^{-9}$,
with which the result is stabilized, and
$K_{\rm max}=20$ and $I_{\rm max}=36$ are used.
The resultant spectrum for the s-band is depicted in Fig.~\ref{fig:Er164gsbr};
for completeness the calculated result of the g-band in Ref.~\cite{STS15}
is also included, which was obtained with the set of five frequencies
$\hbar\omega_{\rm rot}=$0.01, 0.05, 0.10, 0.15, 0.20~MeV,
as in the previous section.
Again, the reference rotational energy, $I(I+1)/80$~MeV, is subtracted.
As it can be seen in the figure,
we have successfully obtained the band crossing between the g- and s-bands,
although the crossing occurs at slightly higher spin 
compared with the observation; at $I\approx 18$ in the calculation,
while between $I=14$ and 16 in the experimental data.
This is non trivial because we have no kind of adjustment
in the present calculation.
It should be emphasized that the g- and s-bands are calculated
independently without the inter-band mixing;
the effect of the inter-band mixing
is not taken into account in the present work.
The reason why the calculated crossing is delayed is that
the calculated excitation energies of the s-band are higher than
the experimental data; the excitation energy of $12^+$ state is 2.69 MeV,
which is about 170 keV higher than the experimentally measured one.

In Figs.~\ref{fig:Er164gb5m} and~\ref{fig:Er164sb4m}, we show
how the resultant spectrum is obtained by the configuration-mixing calculation
for the g- and s-bands, respectively, where the five (four) spectra
calculated by the projection from one cranked HFB state
with different frequencies are depicted in addition to
the result of the configuration-mixing for the g-band (s-band).
From Fig.~\ref{fig:Er164gb5m} it can be seen that
the five spectra for the g-band,
each of which is obtained from the one cranked state
with $\hbar\omega_{\rm rot}=0.01\sim 0.20$ MeV, are rather similar,
and the energy gain by the configuration-mixing is larger at higher spin,
which leads to the increase of the moment of inertia as a function of spin.
This suggests that the $K$-mixing and configuration-mixing
induced by the cranking is more effective at higher spins;
the Hill-Wheeler equation~(\ref{eq:HW}) should be solved
even for the case of projection from a single HFB state
and its dimension of increases as spin increases.
In contrast, the four spectra for the s-band obtained
from the one cranked state with $\hbar\omega_{\rm rot}=0.25\sim 0.40$ MeV
are rather different as is shown in Fig.~\ref{fig:Er164sb4m}.
Non of the spectra looks like the observed one and
the configuration-mixing is very important for the s-band
to obtain the correct spectrum.
Each spectrum in Figs.~\ref{fig:Er164sb4m} has a minimum energy
at a finite spin value,
and the spin value that gives a minimum is larger for the spectrum
obtained from the cranked HFB with larger cranking frequency.
Moreover, the energy gain by the configuration-mixing is considerable
at lower spin, while it is much smaller at $I \gtsim 15$ and
the resultant configuration-mixed spectrum looks more like
the envelope curve of the four spectra.
Note that the aligned angular momentum
of the two quasineutrons is estimated to be about 10 $\hbar$ in $^{164}$Er.
Thus, the role played by the configuration-mixing seems
to be somewhat different in the g- and s-bands,
and the configuration-mixing is much more crucial for the s-band
than for the g-band.

\begin{figure}[!htb]
\begin{center}
\includegraphics[width=90mm]{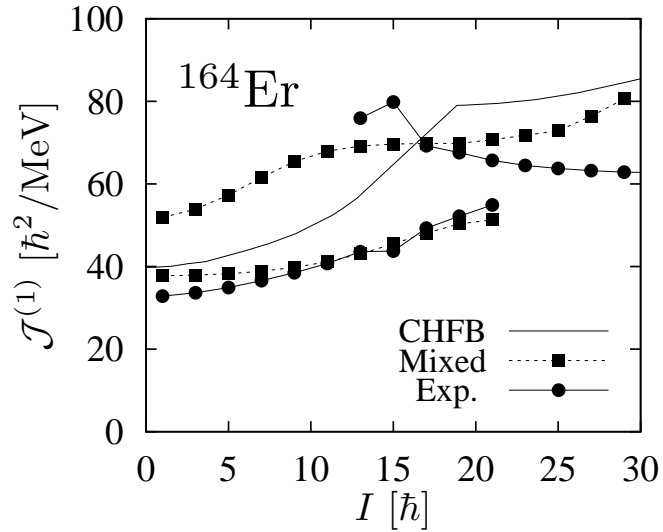}
\vspace*{-4mm}
\caption{
Moments of inertia versus spin value for the g- and s-bands
in $^{164}$Er obtained by the projected configuration-mixing calculations
in comparison with the experimental data.
The result of the cranked HFB (CHFB) is also included.
}
\label{fig:Er164MoIm}
\end{center}
\end{figure}

Although the nice band crossing is obtained by the calculation,
the agreement of the moment of inertia for the s-band is
not as good as that for the g-band, which is shown in Fig.~\ref{fig:Er164MoIm}.
Especially for the s-band, the inertia is considerably overestimated
in $I \gtsim 20$, and the experimentally observed inertia decreases
as spin increases, while the calculated one is almost constant in
the range $10 \ltsim I \ltsim 20$ and increases afterward.
The overestimation of the moment of inertia for the s-band is mainly
due to the fact that the neutron pairing vanishes by the alignment
of two quasineutrons, namely, the cranked HFB states used for
the configuration-mixing calculation for the s-band are
in fact the unpaired states (the Slater determinants) for neutrons.
The proton pairing is non vanishing but
reduces considerably; i.e., the average pairing gap of proton is
$\bar{\Delta}_\pi\approx 0.83$ MeV at $\hbar\omega_{\rm rot}=0.25$ MeV and
$\bar{\Delta}_\pi\approx 0.45$ MeV at $\hbar\omega_{\rm rot}=0.40$ MeV.
This fact is reflected also in the resultant inertia
of the cranked HFB calculation, which is considerably larger at $I \gtsim 20$.
Because of this problem we do not discuss the higher spin part
at $I\gtsim 30$ in the present work.

\begin{figure}[!htb]
\begin{center}
\includegraphics[width=80mm]{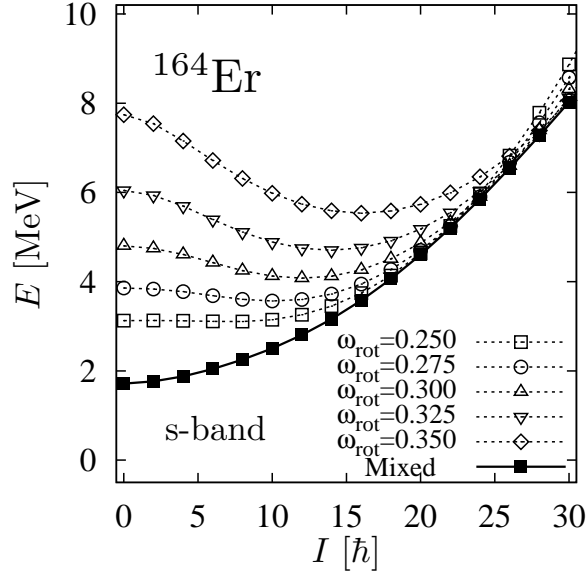}
\vspace*{-4mm}
\caption{
The same as Fig.~\ref{fig:Er164sb4m} but the calculations
with employing five values of the cranking frequencies,
$\hbar\omega_{\rm rot}=$0.250, 0.275, 0.300, 0.325, 0.350~MeV.
}
\label{fig:Er164sb5m}
\end{center}
\end{figure}

\begin{figure}[!htb]
\begin{center}
\includegraphics[width=80mm]{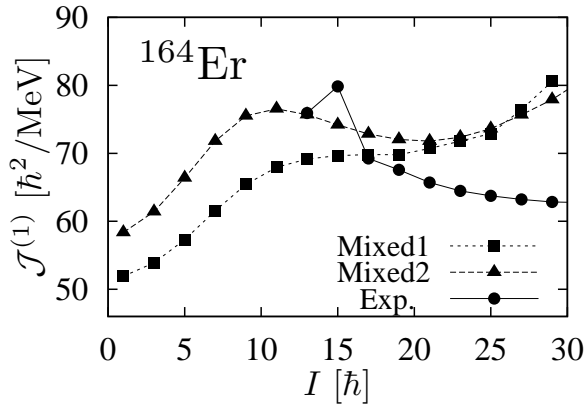}
\vspace*{-4mm}
\caption{
Moments of inertia versus spin value for the s-bands in $^{164}$Er
obtained by the configuration-mixing calculations with
the two different sets of rotational frequencies,
$\hbar\omega_{\rm rot}=$0.25, 0.30, 0.35, 0.40~MeV (Mixed1), and
$\hbar\omega_{\rm rot}=$0.250, 0.275, 0.300, 0.325, 0.350~MeV (Mixed2)
in comparison with the experimental data.
The scale of the ordinate is enlarged from that of Fig.~\ref{fig:Er164MoIm}.
}
\label{fig:Er164MoId}
\end{center}
\end{figure}

In the case of the g-band the result of the configuration-mixing
does not depend on the choice of a set of frequencies for
the cranked HFB states~\cite{STS15}.  However, it turns out that
the result for the s-band is not completely independent of
the choice of frequencies.  We show an example of another set
of the cranking frequencies,
$\hbar\omega_{\rm rot}=$0.250, 0.275, 0.300, 0.325, 0.350~MeV
in Fig.~\ref{fig:Er164sb5m},
where the range of the frequencies are reduced from $[0.25,0.40]$
to $[0.25,0.35]$ and they are chosen more densely in the given interval.
Comparing the resultant configuration-mixed spectra obtained
with the two sets of cranking frequencies
in Figs.~\ref{fig:Er164sb4m} and~\ref{fig:Er164sb5m},
one can see they are slightly different.
To see the difference more clearly,
we show the moments of inertia for the s-band calculated with
these two sets of frequencies in Fig.~\ref{fig:Er164MoId}
in comparison with the experimental data.
Note that the scale of the ordinate is enlarged.
The difference is non-negligible especially
in the lower spin range $I \ltsim 20$.
Moreover, the inertia obtained with the first set,
$\hbar\omega_{\rm rot}=$0.25, 0.30, 0.35, 0.40~MeV (Mixed1),
is monotonically increasing as spin increases,
while the one with the second set,
$\hbar\omega_{\rm rot}=$0.250, 0.275, 0.300, 0.325, 0.350~MeV,
decreases first in $ 10 \ltsim I \ltsim 20$ and then turns to increase.
We think that the result with the second set (Mixed2) is more reliable
in the considered spin-range because a larger number of the cranked HFB states
are employed in the relevant frequency interval.

The expectation values of spin in Eq.~(\ref{eq:IavCHFB})
for the cranked HFB states are $I \approx$ 19.3, 23.6, 29.1, 35.1
at $\hbar\omega_{\rm rot}=$ 0.25, 0.30, 0.35, 0.40~MeV, respectively;
see also Fig.~5 of Ref.~\cite{STS15}.
The cranked HFB state is most suitable for describing the states
with spin around its expectation value.  In fact, the difference of moments of
inertia obtained with the two sets of frequencies is small in the
spin range $I \gtsim 20$; the results of configuration-mixing
mainly differs in the lower spin region, $I \ltsim 10$.
It may be necessary to include the cranked HFB states with lower
spin-expectation values for the configuration-mixing of the s-band
in order to obtain the result that is independent of
the detailed choice of the frequencies.
Thus, the description of the s-band is not as simple as in the case of
the g-band, especially in the low-spin region, $I \ltsim 10$.
It may not, however, be a serious problem because the s-bands
for such low-spin parts have not been observed in experiments and
it is difficult to compare with experimental data.
In any case, we need to study further
for more satisfactory description of the s-band.

\section{Conclusion}
\label{sec:concls}

We have investigated the rotational bands in the rare earth nuclei
by employing our recently developed microscopic framework,
the angular-momentum-projected multi-cranked
configuration-mixing method~\cite{STS15}.
In this method several cranked HFB states are utilized
with a suitably chosen set of rotational frequencies.
We use the Gogny force with the D1S parameter set as an effective interaction,
and there is no ambiguity for the Hamiltonian.

We first apply our method to the g-band of various selected nuclei
in the rare earth region.  Reasonably good overall agreements are obtained
for the energy spectra and the moments of inertia up to about $I \approx 20$.  
In a few cases the moments of inertia at low spin are considerably overestimated
and the increase of the inertia as a function of spin is not enough compared
with the experimental data.  It is found that
the selfconsistently calculated pairing correlations are too weak
for such nuclei; the average pairing gaps for both neutrons and protons
are only about 70\% of or even less than the even-odd mass differences.
If the pairing properties are nicely reproduced,
the agreements of the moments of inertia are found to be excellent.
In this way we have confirmed that our method is capable
to reliably describe the nuclear rotational motion near the ground state.

Next we apply our approach to the study of the s-band in the nucleus $^{164}$Er
for the first time.  The method of calculation is the same for the s-band;
the only difference is that the cranked HFB states
with higher rotational frequencies are employed for the configuration-mixing,
in which the two quasineutrons align their angular momenta.
Thus the g- and s-bands can be calculated separately
without the inter-band mixing between them.
The band crossing between the g- and s-bands can be reproduced,
although the spin value, at which the two bands cross, is slightly larger
than the observed one.
The calculated moment of inertia of the s-band is overestimated
especially at high-spin states.  This is mainly because the selfconsistently
calculated pairing correlation for neutrons vanishes in the cranked HFB states
due to the alignment of two quasineutrons,
which cannot be avoided as long as the Gogny D1S force is employed.
It is found that the result of configuration-mixing weakly depends on
the choice of the set of cranking frequencies for the s-band,
especially in the lower spin region, in contrast to the case of the g-band,
where the result is independent of the choice of frequencies.
Thus, further investigation is necessary for
the proper description of the s-band, which is an important future work.

\section*{ACKNOWLEDGEMENTS}

This work is supported in part
by Grant-in-Aid for Scientific Research (C) 
No.~25$\cdot$949 from Japan Society for the Promotion of Science.

\vspace*{10mm}



\begin{thebibliography}{99}

\bibitem{BM75}
A.~Bohr and B.~R.~Mottelson,
{\it Nuclear Structure}, Vol.~II Benjamin, New York (1975).

\bibitem{BF79}
R.~Bengtsson and S.~Frauendorf,
Nucl.\ Phys.\ A \textbf{327}, 139 (1979).
 
\bibitem{BFM86}
R.~Bengtsson, S.~Frauendorf, and F.-R.~May,
Atomic Data and Nuclear Data Table  \textbf{35}, 15 (1986).

\bibitem{RS80}
P.~Ring and P.~Schuck,
{\it The Nuclear Many-Body Problem}, Springer, New York (1980).
 
\bibitem{TS12}
S.~Tagami and Y.~R.~Shimizu,
Prog.\ Theor.\ Phys.\ \textbf{127}, 79 (2012).

\bibitem{TS16}
S.~Tagami and Y.~R.~Shimizu,
Phys.\ Rev.\ C \textbf{93}, 024323 (2016).

\bibitem{STS15}
M.~Shimada, S.~Tagami, and Y.~R.~Shimizu,
Prog.\ Theor.\ Exp.\ Phys.\ \textbf{2015}, 063D02 (2015).

\bibitem{PT62}
R.~E.~Peierls and D.~J.~Thouless,
Nucl.\ Phys.\ \textbf{38}, 154 (1962).

\bibitem{RER15}
M. Borrajo, T. R. Rodr\'iguez, J. L. Egido,
Phys.\ Lett.\ B \textbf{746}, 341 (2015).

\bibitem{HS95}
K.~Hara and Y.~Sun,
Int.\ J.\ Mod.\ Phys.\ E \textbf{04}, 637 (1995).

\bibitem{DeGo80}
J.~Decharg\'e and D.~Gogny,
Phys.\ Rev.\ C \textbf{21}, 1568 (1980).

\bibitem{D1S}
J.~F.~Berger, M.~Girod, and D.~Gogny,
Comput.\ Phys.\ Commun.\ \textbf{63}, 365 (1991).

\bibitem{IMYM02}
T.~Inakura, S.~Mizutori, M.~Yamagami, and K.~Matsuyanagi,
Nucl.\ Phys.\ A \textbf{710}, 261 (2002).

\bibitem{AW03}
G.~Audi, A.~H.~Wapstra, and C.~Thibault,
Nucl.\ Phys.\ A \textbf{729}, 337 (2003).

\bibitem{Shimada16}
M.~Shimada S.~Watanabe, S.~Tagami, T.~Matsumoto, Y.~R.~Shimizu,
and M.~Yahiro, to be published.

\bibitem{NNDC}
Brookhaven database, http:/\!/www.nndc.bnl.gov.

\bibitem{TOI}
Table of Isotope homepage, http:/\!/ie.lbl.gov/toi.html.

\bibitem{JRH72}
A.~Johnson, H.~Ryde, and S.~Hjorth,
Nucl.\ Phys.\ A \textbf{179}, 753 (1972).

\end{thebibliography}
\end{document}